\documentclass{article}

\usepackage{amsfonts}
\usepackage{amstext}
\usepackage{color}
\usepackage{amsthm}
\theoremstyle{plain}
\definecolor{Red}{cmyk}{0,1,1,0}

\definecolor{Blue}{cmyk}{1,1,0,0}

\definecolor{Pink}{cmyk}{0,1,0,0}

\definecolor{Green}{cmyk}{1,0,1,0.5}

\newcommand{\ba}{\begin{array}}
	\newcommand{\ea}{\end{array}}
\newcommand{\be}{\begin{equation}}
\newcommand{\ee}{\end{equation}}
\newcommand{\ben}{\begin{enumerate}}
	\newcommand{\een}{\end{enumerate}}

\newcommand{\eop}{\hfill \rule{0.7ex}{1.6ex}}

\newcommand{\R}{\mathbb{R}}

\newcommand{\B}{{\cal B}}
\newcommand{\Z}{\mathbb{Z}}

\newtheorem{teo}{Theorem}[section]
\newtheorem{lema}{Lemma}[section]
\newtheorem{cor}{Corollary}[section]
\newtheorem{prop}{Proposition}[section]

\title{
	{\large{ \bf{HIGHER ORDER ASYMPTOTICS OF DECAYING SOLUTIONS OF SOME GENERALIZED BURGERS EQUATIONS}}}}

\date{}

\begin{document}
	\maketitle

	\centerline{\scshape Gast\~ao A. Braga}
	\medskip
	{\footnotesize
		\centerline{Departamento de Matem\'atica}
		\centerline{Universidade Federal de Minas Gerais}
		\centerline{Caixa Postal 1621, Belo Horizonte, 30161-970, Brazil}
	} 
	
	\medskip
	
	\centerline{\scshape Frederico Furtado}
	\medskip
	{\footnotesize
		\centerline{Department of Mathematics and Statistics}
		\centerline{University of Wyoming}
		\centerline{E. University Ave. Laramie, WY 82071, U.S.A.}
	} 
	
	\medskip
	
	\centerline{\scshape Jussara M. Moreira}
	\medskip
	{\footnotesize
		\centerline{Departamento de Matem\'atica}
		\centerline{Universidade Federal de Minas Gerais}
		\centerline{Caixa Postal 1621, Belo Horizonte, 30161-970, Brazil}
	}
	
	\medskip
	
	\centerline{\scshape Ant\^onio Marcos da Silva}
	\medskip
	{\footnotesize
		\centerline{Departamento de Matem\'atica}
		\centerline{Universidade Federal de Ouro Preto}
		\centerline{R. Diogo de Vasconcelos, 122, Pilar, 35400-000, Brazil}
	}

	\def\l{\lambda}
	
	\baselineskip = 22pt
	
	\maketitle
	
	\begin{abstract}
	We study the large-time behavior of solutions to a generalized Burgers Equation, with initial zero mass data. Our main purpose is to present a modified version of the Renormalization Group map, which is able to provide the higher order asymptotic properties of the solution to the Cauchy problem of a class of nonlinear time-evolution problems.
	\end{abstract}
	\clearpage

	\section{\large{Introduction}}
	\label{sec:intr}
	
The Renormalization Group (RG) Methods have been applied to numerous problems since the early 1950s. Their application
to the asymptotic analysis of differential equations was initiated in the 1990s \cite{bib:gold-book, bib:gold2, bib:gold1, bib:oono} and the mathematical aspects of the method were rigorously developed by Bricmont, Kupiainen and collaborators \cite{bib:bric-kupa-lin}. The later has shown to be very efficient in providing the leading
order long time asymptotics of solutions to a wide class of initial value problems, both analytically and numerically \cite{bib:cag, bib:bona, bib:merd, bib:vin}.
This is so as long as we consider nonzero mass initial data. Under the zero mass condition, the
scaling used in \cite{bib:bric-kupa-lin} to define the RG operator
fails to provide a method good enough to describe refined asymptotic information such
as profile function and decay or spreading exponents.
Despite this, the zero mass condition is crucial if one wants to look for higher order asymptotics.
Also, in many interesting problems, as for instance the extension to the whole line of the
mixed initial value problem (IVP) on a half line with Dirichlet zero boundary condition at the origin,
the zero mass condition naturally appears.

In this paper we extend the RG method as developed in \cite{bib:bric-kupa-lin}
to IVPs under the zero mass condition on the initial data.
We show that, even if $\int_\R f(x)dx = 0$, a suitable scaling can be conveniently
defined, which generates an RG operator whose iterates converge to a fixed point that correctly describes the profile
function and the decaying exponents associated with the next-to-leading order asymptotics of solutions to the IVP.

More specifically, we deal with the following class of problems:
	\begin{equation}
		\left\{
		\begin{array}{ccccccc}
			u_t +  uu_x= u_{xx} + \lambda F(u,u_x), \  x \in \mathbb{R}, t > 1, \\
			u(x, 1) = f(x).
			\label{pvi:n-lin-F}
		\end{array}
		\right.
	\end{equation}
	For small enough initial data and if one of the two hypothesis is satisfied: \newline
 {\bf{(H-1)}} $\int_\R f(x)dx = 0$ and $F(u,u_x) = \displaystyle\sum_{n\geq 2}c_nu^nu_x$, \newline
{\bf{(H-2)}} $f$ is odd 
and
		$		F(u,u_x) =  \displaystyle\sum_{m\geq a,\, n \geq b \atop 4a+3b-2>0}c_{m,n} u^{2m+1}u_x^{n}$, \newline
we will show that the solution $u(x,t)$ to IVP (\ref{pvi:n-lin-F}) behaves, for $t \gg 1$, as
		\begin{equation}
			u(x,t) \approx \frac{A}{t}f_1^*\left(\frac{x}{\sqrt{t}}\right),
			\label{eq:comp-assi-pvi}
		\end{equation}
		where $A$ is a prefactor and
		\begin{equation}
			f_1^*\left(x\right) =
			-\frac{x}{2}
			\frac{e^{-\frac{x^2}{4}}}{\sqrt{4\pi}}.
			\label{def:pont-fixo}
		\end{equation}
		The long time behavior (\ref{eq:comp-assi-pvi}) will come out from the iterates
		of a nonlinear operator (the RG operator) whose linearization will have  $f_1^*\left(x\right)$,
		given by (\ref{def:pont-fixo}), as a fixed point. Furthermore, the time decay exponent $\alpha = 1$ and the
		time spread exponent $\beta = 1/2$, on the right hand side of (\ref{eq:comp-assi-pvi}),
		are intimately related to the definition of the RG operator.
				The nonlinearity $F(u, u_x)$ in  {\bf{(H-1)}} or  {\bf{(H-2)}} is such that it preserves the
		symmetry of the initial data along the time evolution. Also,
		$F(u, u_x)$ is chosen to be ``irrelevant'' under the RG flow so that the long time behavior
		(\ref{eq:comp-assi-pvi}) will be, essentially, the one given by the linearized problem.
		
		The motivation for this paper arose from our attempt to understand, within the context of the
	RG framework as developed by Bricmont et al.
	in \cite{bib:bric-kupa-lin}, those numerical results presented
	by Braga et al. in \cite{bib:brag-Furt-isa-Lee} regarding
	the Burgers Equation under the zero mass condition.
	This is exactly IVP (\ref{pvi:n-lin-F}), under {\bf{(H-1)}} and  $\lambda = 0$,
	and the  Numerical Renormalization Group
	(NRG) scheme developed in \cite{bib:brag-Furt-isa-Lee} was able to predict,
	within the allowed numerical error, the long time behavior (\ref{eq:comp-assi-pvi}),
	the profile function being the expected one (\ref{def:pont-fixo})
	as well as the time exponents $\alpha=1$ and $\beta=1/2$,
	being this a strong evidence that it would be possible to analytically establish a multiscale
	argument to provide the asymptotic behavior
	by first principles.
	 The correctness of the nRG results is validated by Whitham in \cite{bib:whitham},
	where an explicit formula for the solution to (\ref{pvi:n-lin-F}) with $\lambda=0$ is obtained.
A $t\to\infty$ analysis of this formula validates the behavior predicted numerically in \cite{bib:brag-Furt-isa-Lee}. 
	On the other hand, a straightforward application of the multiscale analysis of
	\cite{bib:bric-kupa-lin} to this problem can only lead to the conclusion
	that $u (x, t) \approx 0 $ if $\int_\R f(x) dx = 0$ and $ t \gg 1 $.
	Therefore, if one wants to preserve the ideias behind the
	multiscale method of \cite{bib:bric-kupa-lin}
	to get the long time behavior (\ref{eq:comp-assi-pvi}),
	it is necessary to modify the definition of the RG operator to incorporate
	the correct time exponents $\alpha$ and $\beta$ and that is what we have
	done in this paper.
	
We point out that Bona et al., in \cite{bib:bona},  have used the RG approach to
prove that, at lowest order, the asymptotic state of solutions to a generalized
Kortweg-de Vries equation do not depend upon the nonlinearity, the dispersion,
nor on the initial data (except for its mass). In order to visualize the effects
of nonlinearity and dispersion, the authors performed in \cite{bib:bona1}
a higher order asymptotic analysis, but without changing the usual definition of the
RG operator, which demanded a greater refinement in the performed calculations.
	
 For $t \gg 1$, the behavior (\ref{eq:comp-assi-pvi}) can be rephrased as
$tu(\sqrt{t} x, t) \approx A f_1^*\left(x\right) $ and taking $t= L^2$, $L> 1$ ,
we are lead to the scaling $ L^2u (Lx, L^2)$
(instead of the canonical rescaling $Lu(Lx, L^2)$, see \cite{bib:bric-kupa-lin}). This being the case,
terms of the form $ u ^ a u_x ^ b $, in $ u_t = u_ {xx} + u ^ a u_x ^ b $,
would be contracted by a factor of $ L^{2a + 3b - 4}$ if $ 2a + 3b - 4< 0 $.
In particular, this would be the case for the Burgers $uu_x$ term as well as
for the nonlinearity $F(u, u_x)$ assumed in {\bf{(H-1)}} and {\bf{(H-2)}},
that is, they will all be ``irrelevant'' in the limit $ L \gg 1 $.
Notice that, with the canonical rescaling as in \cite{bib:bric-kupa-lin}, the nonlinearity $uu_x$ is {\it marginal}, i.e.,
the equation $u_t=u_{xx} + \lambda  uu_x$ is scale invariant.

The asymptotic behavior (\ref{eq:comp-assi-pvi}) can also be
heuristically evidenced: if we consider the explicit
solution of the linear problem given by an inverse Fourier transform, it is clear that
\begin{equation}
\label{eq:comp-gaus}
u(x,t) \approx \frac{\hat{f}(0)}{\sqrt t}\phi\left(\frac{x}{\sqrt t}\right), \,\,\, t\gg 1,
\end{equation}
where $\phi$ is the Gaussian distribution.
However, if one considers a zero mass initial data, then
(\ref{eq:comp-gaus}) reduces to $ u (x, t) \approx 0 $, giving us no
quantitative information about how $ u (x, t) $ approaches zero.
In this case, the above approach becomes
\begin{eqnarray}
\label{aprox-formal}
	u(x,t) \nonumber
	& \approx &
\frac{-i\hat{f}'(0)}{2 \pi t}
	\int_{\R}i\omega e^{-\omega^2} e^{i\omega \left(\frac{x}{\sqrt{t}}\right)}d \omega
	+ e^{-t}O\left(t^{-1}\right),
	\,\,\, t \gg 1,
	\\
	& \approx &
\frac{A}{t}f_1^*\left(\frac{x}{\sqrt{t}}\right),
\ t \gg 1,
\end{eqnarray}
where $A = -i\hat{f}'(0)=-\int_{\R}xf(x)dx$ and $f_1^*$ is given by (\ref{def:pont-fixo}).
Notice that the behavior (\ref{aprox-formal}) is similar to (\ref{eq:comp-assi-pvi}).

We now state our results. Since we are now considering $f$ under the zero mass condition, it is
necessary to incorporate the second derivative term in the definition of the space
	for the initial data. More precisely, given $q > 1$, define
	\begin{equation}
		\|f\|_q = \sup_{\omega \in \R}(1+|\omega|^q)
		\left(|\widehat{f}(\omega)|+|\widehat{f}'(\omega)|+|\widehat{f}''(\omega)|\right)
		\label{def:norm-bq}
	\end{equation}
	and
	\begin{equation}
		\label{def:esp-bq}
		\B_q=\left\{f\in L^1(\R): \widehat{f}(\omega) \in C^{2}(\R) \ \mbox{e} \ \|f\|_q<+ \infty \right\}.
	\end{equation}

In Section \ref{sec:arg} we validate the results obtained numerically in \cite{bib:brag-Furt-isa-Lee} by proving the following theorem:
			\begin{teo}
				\label{teo:lim-ass-burgers}
				For fixed $q>3/2$, let $f\in \B_q$ satisfying the zero mass condition and consider IVP (\ref{pvi:n-lin-F}) with $\lambda = 0$.
				There are $\epsilon > 0$ and $A = A(f, uu_x)$ such that, if $\|f\|_q<\epsilon$, then the solution $u$ to IVP (\ref{pvi:n-lin-F})
		satisfies
				\begin{equation}
					\label{lim-ass-burgers}
					\lim_{t \longrightarrow + \infty}
					\|tu(t^{\frac{1}{2}}\cdot, t) - Af_1^*\|_q	= 0,
				\end{equation}
				where $f_1^*$ is given by (\ref{def:pont-fixo}).
			\end{teo}
		
			{\bf Remark:} Although we have assumed $\lambda = 0$, the above	theorem also holds
			 for $\lambda \not = 0$ and $F(u, u_x)$ given in {\bf{(H-1)}}) and its proof goes along the
			lines of Section \ref{sec:arg} together with the additional care to guarantee that iterates will
			all end up inside the power series' interval of convergence (see also Section \ref{sec:geral}).
			
			In Section \ref{sec:geral} we consider a more general problem and prove our main result: 
			\begin{teo}
				\label{teo:lim-ass-n-lin}
				Given $q>2$, consider IVP (\ref{pvi:n-lin-F}) satisfying hypothesis {\bf{(H-2)}}, $f\in \B_q$ and $|\lambda|\leq 1$.
                There are $\epsilon > 0$ and $\bar{A}=\bar A(f, F)$ such that, if $\|f\|_q<\epsilon$, then
				the solution $u$ to IVP (\ref{pvi:n-lin-F}) satisfies
				\begin{equation}
					\label{lim-ass-n-lin}
					\lim_{t \longrightarrow + \infty} \|tu(t^{\frac{1}{2}}\cdot, t) - \bar{A}f_1^*\|_q = 0,
				\end{equation}
				where $f_1^*$ is given by (\ref{def:pont-fixo}).
			\end{teo}

As a corollary to the above theorem, we obtain the asymptotics of the
solution for a mixed Dirichlet problem on the half-line. Consider the IVP
\begin{equation}
		\left\{
		\begin{array}{ccccccc}
			u_t +  uu_x= u_{xx} + \lambda F(u,u_x), \  x>0, t > 1, \\
			u(x, 1) = f(x), \,\,\, u(0, 1) =0.
			\label{pvi:n-lin-F-half-line}
		\end{array}
		\right.
	\end{equation}
\begin{cor}
	\label{cor:probl-dirichlet}
	Under the hypothesis {\bf{(H-2)}}, there are $\epsilon > 0$
	and $\bar{A}=\bar A(f, F)$ such that, if $\|\tilde{f}\|_q<\epsilon$ where
	$\tilde{f}$ is the odd extension of $f$, then the solution $u$ of (\ref{pvi:n-lin-F-half-line})
	satisfies
	\begin{eqnarray*}
	\lim_{t \longrightarrow + \infty} \|tu(t^{\frac{1}{2}}\cdot, t) - \bar{A}f_1^*\|_q = 0,
	\end{eqnarray*}
	where $f_1^*$ is given by (\ref{def:pont-fixo}).
\end{cor}

\section{The Algorithm and the Burgers Equation}
\label{sec:arg}

The numerical results and the nRG generator algorithm in
\cite{bib:brag-Furt-isa-Lee} are strong indications that the Renormalization Group method could be implemented
analytically to study the initial value problem associated with the Burgers Equation.
In this section we will show that the numerical result corresponds to the dynamics of the RG operator, in the vicinity of its critical point.
We will see that the long time behavior of the solution to IVP (\ref{pvi:n-lin-F}) with $\lambda = 0$ and zero mass initial data is the
same as the one corresponding to the linear IVP, that is, the term $uu_x$ does not contribute to the asymptotic behavior,
except for the prefactor.

To proceed, we first need to prove that, given $L> 1$, $q> \frac{3}{2}$ and $\lambda \in [-1,1] $, if the initial data $f$
is
sufficiently ``small'' in $\B_q$, then, the IVP
\begin{equation}
\label{pvi:Eq-de-Burg}
\left\{
\begin{array}{cccccc}
u_t & = &u_{xx}+\lambda uu_x,
\,\,\, x\in \R, \ t \in (1, L^2], \\
u(x,1) & = & f(x),
\,\,\, f \in \B_q.
\end{array}
\right.
\end{equation}
with $\lambda \in [-1,1]$, has a unique local solution with $t \in [1, L ^ 2]$.
Consider
\begin{equation}
	\label{def:esp-B}
B^{(L)}= \left\{u:\mathbb{R} \times [1,L^2]\longrightarrow \mathbb{R}; \ u(\cdot, t) \in B_q, \ \forall t \in [1,L^2]
\right\},
\end{equation}
\begin{equation}
\label{def:norm-B}
\|u\| = \sup_{t \in [1,L^2]}
\|u(\cdot, t)\|_q.
\end{equation}
and let $u_f$ be the solution to the linear equation associated with the Burgers equation. Then, defining the ball
\begin{equation}
\label{def:bola-Bf}
B_f = \left\{u \in B/ \|u - u_f\| \leq \|f\|_q\right\}
\end{equation}
we have the following
\begin{teo}
	\label{teo:exis-uni}
	Given $L>1$, $q>\frac{3}{2}$ and $\lambda \in [-1,1]$, there exists $\epsilon = \epsilon (L,q) > 0$ such that,
	if $f \in \B_q$ and $\|f\|_q< \epsilon$, then, IVP (\ref{pvi:Eq-de-Burg}) has a unique solution in $B_f$.
\end{teo}
Defining the operator
$T: B^{(L)} \longrightarrow B^{(L)}$, $u  \mapsto  u_f + \lambda N(u)$,
where
\begin{equation}
N(u)(x,t) = \int_0^{t-1} \int_{\mathbb{R}} \frac{e^{\frac{-(x-y)^2}{4s}}}{\sqrt{4 \pi s}}\cdot \frac{(u^2)_x}{2}(y,t-s-1)dy ds,
\label{def:operador-N}
\end{equation}
the proof of the above theorem is straightforward from the Banach Fixed Point Theorem once it is shown that the operator $T$ is such that $T(B_f)\subset B_f$ and it is a contraction in $B_f$. In Section \ref{sec:geral} we give a more detailed proof of Theorem \ref{teo-ex-uni-caso-geral}, which generalizes the above local existence and uniqueness theorem.

The RG approach that we employ in this paper is basically the integration of the equation followed by a rescaling. To explain this idea, we
let $u$ be a real-valued function of $(x, t) \in \R\times R^+$. For a fixed $L > 1$, define, inductively, a sequence of rescaled functions $\{u_n\}_{n=0}^{\infty}$,
by $u_0=u$ and, for $n\geq 1$,
$$
u_n(x,t)=L^2u_{n-1}(Lx, L^2t).
$$
If the original function $u$ is a global solution to IVP (\ref{pvi:Eq-de-Burg}), then a direct calculation reveals
that $u_n$ satisfies the renormalized IVP:
\begin{equation}
\label{pvi:eq-de-Burg-n}
\left\{
\begin{array}{cccccc}
u_t & = &u_{xx}+\lambda_n uu_x,
\,\,\, x\in \R, \,\,\, t \in (1, L^2], \\
u(x,1) & = & f_n (x),
\,\,\, f_n \in \B_q, \,\,\, q > \frac{3}{2},
\end{array}
\right.
\end{equation}
where $\lambda_n= \lambda_0 L^{-n}$, with $\lambda_0=\lambda$ and $f_0=f$.
From Theorem \ref{teo:exis-uni}, if $\|f_n\|<\epsilon$, there is a unique solution to IVP (\ref{pvi:eq-de-Burg-n}) in $B_{f_n}$, which can be written as
$u_n(x,t)=u_{f_n}(x,t)+\nu_n(x,t)$,
where $u_{f_n}$ is the solution to the linear problem with initial data $f_n$ and $\nu_n = \lambda_nN(u_n)$, with $N$ given by (\ref{def:operador-N}).
We can then define the RG operator for IVP (\ref{pvi:eq-de-Burg-n}) with $n \geq 0$:
\begin{equation}
\label{def:RG-n-lin}
\left(R_{L,n}f_n\right)(x) \equiv
L^2u_n(Lx, L^2), \ \forall x \in \R
\end{equation}
and
\begin{equation}
f_{n+1} \equiv R_{L,n}f_n.
\label{eq:dado-ini-fn-n-lin}
\end{equation}
With the above definitions in mind, we now introduce the steps to construct the iterative process behind the RG method.
Define
\begin{equation}
\label{eq:bola-epsi}
B_\epsilon =\{f\in B_q: \|f\|_q<\epsilon\},
\end{equation}
where $\epsilon>0$ is given by Theorem \ref{teo:exis-uni}.
We will call the following procedure the {\em Renormalization Group generator Algorithm}, or simply RGA.
Start with the initial condition of the IVP (\ref{pvi:Eq-de-Burg}), $f_0 = f\in B_\epsilon$, under the zero mass condition.
For $n = 0, 1, 2, . . .$, we have the following:
\begin{enumerate}
\item $f_n$ is decomposed in two components, one of them in the direction of the fixed point $f_1^*$.
The other, given by $f_n - A_nf_1^* = g_n$, depends on the choice of the prefactor $A_n$. The later is then chosen in such a way that
$\widehat{g_n}(0)=\widehat{g_n}'(0)=0$, that is, $g_n$ will be contracted by the RG operator (see Lemma \ref{lema:contract});
\item if $f_n\in B_\epsilon$, then the IVP (\ref{pvi:eq-de-Burg-n}) has a unique solution $u_n$ in $B_f$ and, therefore, $L^2u_n(Lx, L^2)$
is well defined and it is in $\B_q$;
\item $L^2u_n(Lx, L^2)$ defines a new initial data $f_{n+1}$ for a new, renormalized PDE, which 
differs from the previous one by the parameter $\lambda_{n+1} = \lambda_n L^{-1}$.
From the symmetry of the equation, $f_{n+1}$ has zero mass if $f_n$ does.
\item estimate $\|f_{n+1}\|_q$ to guarantee that $f_{n+1} \in B_\epsilon$ if $f\in B_\epsilon$ is well chosen, in order to iterate the process.
\item\label{step5} estimate the distance between $A_{n+1}$ and $A_n$ and how big is $g_{n+1}$ when compared to $g_n$ and $f_n$.
\end{enumerate}
Assuming that the ARG can be iterated, we will have a sequence of initial value problems, with zero mass $ f_n $
and parameter $ \lambda_n = \lambda L^{-n} $. In addition, $ f_n $
will admit the desired decomposition $A_nf_1^*+g_n$ and will be in the $ B_\epsilon$ ball. Step \ref{step5} above will allow us to prove that,
when $n$ goes to $\infty$, $g_n$ converges to zero and $A_n$ converges to some value $A$ so that $f_n \to A f_1^*$.
In order for us to state and prove the {\em{renormalization lemma}}, which will enable the iteration of the algorithm described above, we need to define the linear operator and state some of its important properties.

Let $u$ be the solution to IVP (\ref{pvi:Eq-de-Burg}) with $\lambda=0$ and define the {\it linear RG operator} $R_L:B_q \to B_q$,
by $(R_Lf)(x) \equiv L^2u(Lx, L^2)$.
It is not hard to see that
\begin{eqnarray}
\label{des:cota-pont-fixo}
\|f_1^*\|_q \leq \sup_{\omega \in \R}(1+|\omega|^q)\left(1 + 7|\omega| + 2 \omega^2 + 4|\omega^3|\right)e^{-\omega^2}\equiv k_q < \infty,
\end{eqnarray}
which implies that $f_1^* \in B_q$. It then follows that $f_1^*$ is a fixed point of $R_L$, i.e., $R_Lf_1^*=f_1^*$.
Furthermore, the semigroup property $R_L \circ \cdots \circ R_L= R_{L^n}$, for $n-1$ compositions of $R_L$, holds.
Finally, one important property of the linear RG operator, essential for determining the
asymptotic behavior, not only in the linear but also in the nonlinear cases, is the fact that the operator contracts functions with both zero mass and zero first moment:
\begin{lema}[Contraction Lemma]
\label{lema:contract}
Given $L>1$ and $q>1$, for $g \in \B_q$ satisfying $\widehat{g}(0)=\widehat{g}'(0)=0$, there are $C=C(q)>0$ and $L_0>1$ such that
$$
\|R_Lg\|_q \leq \frac{C}{L}\|g\|_q,
$$
for all $L >L_0$.
\end{lema}
The proof of the above theorem uses basically the Fundamental Theorem of Calculus, the definition of the $B_q$ space and the properties of the Fourier Transform.
We are finally able to state the
\begin{lema}
	\label{lema:renormal}[Renormalization Lemma]
	Given $L>L_0$, consider IVP (\ref{pvi:eq-de-Burg-n}) with initial data $f_n \in B_\epsilon$ with zero mass and such that
	\begin{equation}
	f_n=A_nf_1^*+g_n,
	\label{eq:decomp-fn}
	\end{equation}
	where $A_n$ is a constant, $f_1^*$ is a fixed point of the linear RG operator, see (\ref{def:pont-fixo}), and
 $g_n \in B_q$, $\widehat{g_n}(0)=\widehat{g_n}'(0)=0$. Then:
	\begin{enumerate}
		\item[ (a)] $f_{n+1}$ given by (\ref{eq:dado-ini-fn-n-lin}) admits the decomposition
		$f_{n+1}=A_{n+1}f_1^*+g_{n+1}$, where
		$A_{n+1}= 	A_n-i\widehat{\nu_n}'(0)$ and
		$g_{n+1} = R_Lg_n + L^2\nu_n(L\cdot) +i\widehat{\nu_n}'(0)f^*_1$.
		Furthermore, $g_{n+1}\in B_q$ is such that
		$\widehat{g_{n+1}}(0)=\widehat{g_{n+1}}'(0)=0$.
		In particular, $f_{n+1}$ has zero mass;
		\item[ (b)] There are constants $G_{L,q}$ and $E_{L,q}$, depending on $L$ and $q$ such that, $|A_{n+1}- A_n| \leq |\lambda_n|G_{L,q}\|f_n\|_q^2$ and
$\|g_{n+1}\|_q \leq \frac{C}{L}\|g_n\|_q+|\lambda_n|E_{L,q}\|f_n\|_q^2$,
where $C$ is the constant in the Contraction Lemma \ref{lema:contract}.
	\end{enumerate}
\end{lema}
\proofname: The proof follows closely the one in \cite{bib:braga-furt-mor-rolla-tp}. Notice that, in this case, since
$\|f_n\|_q<\epsilon$, it follows from (\ref{def:RG-n-lin}),
(\ref{eq:dado-ini-fn-n-lin}) and from decomposition (\ref{eq:decomp-fn}), that
$f_{n+1} (x) = A_nf_1^*(x) + R_Lg_n(x) + L^2\nu_n(Lx)$, where $\nu_n(L \cdot) = \nu_n(L\cdot, L^2)$, from which we get item (a). In particular,
since $\widehat{g_{n}}(0)=\widehat{f_1^*}(0)= \widehat{\nu_n}(0) = 0$, we get that $f_{n+1}$ has zero mass.
Item (b) follows from the fact that
\begin{equation}
\label{des:oper-N}
\|u_n\| \leq \bar{C}_L\|f_n\|_q, \,\, {\mbox{ and }} \|N(u_n)\| \leq G_{L,q}\|f_n\|^2_q,
\end{equation}
\begin{equation}
\bar{C}_L = \bar{C}(L) \equiv 6L^2+4\sqrt{L^2-1}-4.
\label{eq:C-barra}
\end{equation}
and
$$
G_{L,q} \equiv \frac{\bar{C_L}^2(2^{q+1}+3)}{12\pi}(4L^6+6L^4-6L^2+137)\int_{\R}\frac{1}{1+|x|^q}dx.
$$
Hence,
$
|A_{n+1} - A_n| \leq |\widehat{\nu_n}'(0)| \leq
|\lambda_n|G_{L,q}\|f_{n}\|_q^2
$
and also, from the decomposition of $g_{n+1}$, the Contraction Lemma and (\ref{des:oper-N}),
$\|g_{n+1}\| \leq
CL^{-1}\|g_n\|_q+|\lambda_n|E_{L,q}\|f_n\|_q^2$,
where $E_{L,q}=G_{L,q}(L^{q+1} + k_q)$, with $k_q$ given by (\ref{des:cota-pont-fixo}).
\eop

For $\delta \in (0,1)$, define
\begin{equation}
\label{eq:L-delta}
L_{\delta} = L(\delta, q)
\equiv
\max\{L_0,[2C(1+k_q)]^{\frac{1}{\delta}}\},
\end{equation}
where $L_0$ and $C$ are given in the Contraction Lemma \ref{lema:contract} and $k_q$ is given by (\ref{des:cota-pont-fixo}).
From now on, assume that $L>L_\delta$. Define
\begin{equation}
\label{D1geral}
D_1 =
\frac{1}{L^{1-\delta}}
+
k_q
\left(1+G_{L,q}\|f_0\|_q\right)
\end{equation}
and, for $k \in \{1, 2, \cdots \}$,
\begin{eqnarray}
\label{def:D-k}
 D_{k+1} =
 \frac{1}{L^{(k+1)(1-\delta)}}
 +
 k_q\left(1+G_{L,q}\|f_0\|_q+
 G_{L,q}\|f_0\|_q
 \sum_{j=1}^{k}
 \frac{D_j^2}{L^{j}}
 \right).
\end{eqnarray}
Let $D$ be
\begin{equation}
D \equiv 1+ k_q\sum_{j=0}^{+\infty}\frac{1}{L^{j(1-\delta)}},
\label{eq:D}
\end{equation}
where $G_{L,q}$ is the constant in item (b) of Lemma \ref{lema:renormal}.
Notice that, if $\|f_0\|_q<(2L^{1-\delta}{E}_{L,q}D^2)^{-1}$,
we can show that $D_k<D$, for all $k \in \Z^+$, with $D$ and $D_k$ given, respectively by (\ref{eq:D}) and (\ref{def:D-k}).
If $\epsilon>0$ is the one given in Theorem \ref{teo:exis-uni} and the initial data $f=f_0$ is such that $\|f_0\|_q<\epsilon$, then we can iterate the ARG algorithm as long as we can guarantee that $\|f_n\|_q<\epsilon$. This condition is fullfilled if $\|f_0\|_q$ is suficiently small as the next result shows.

\begin{teo}
	\label{teo:cota-fn-gn-An}
Given $\delta \in (0,1)$ and $L>L_{\delta}$, 
there exists $\bar{\epsilon}>0$ such that, if $\|f_0\|_q<\bar{\epsilon}$ and if $f_0$ has zero mass, then,
for all $n=1, 2, \cdots$, $f_n$ given by (\ref{eq:dado-ini-fn-n-lin}) is well defined, has zero mass and admits representation
(\ref{eq:decomp-fn}), where $\hat g_n(0)=\hat g_n'(0)=0$ and
\begin{equation}
\|g_n\|_q
\leq
\frac{1}{L^{n(1-\delta)}}\|f_0\|_q.
\label{des:decaim-gn}
\end{equation}
Furthermore,
\begin{equation}
\|f_n\|_q \leq D_n\|f_0\|_q,
\label{des:cota-fn}
\end{equation}
with $D_n$ given by (\ref{def:D-k}) and, in particular, $\|f_n\|_q < \epsilon$.
\end{teo}
\proofname:
Define
\begin{equation}
\label{eq:epsilon-barra}
\bar{\epsilon} \equiv
\min
\left\{\frac{1}{2L^{(1-\delta)}E_{L,q}D^2},
\frac{\epsilon}{D}
\right\},
\end{equation}
where $\epsilon>0$ is given by Theorem \ref{teo:exis-uni},
$E_{L,q}$ given in Lemma \ref{lema:renormal} and $D$ given by (\ref{eq:D}). Then, the proof follows from induction on $n$ together with the application of ARG algorithm.
\eop

It follows from Lemma \ref{lema:renormal} and Theorem \ref{teo:cota-fn-gn-An} that $(A_n)$ is a convergent sequence. In fact, we have the following:
\begin{cor}
	\label{cor:An-cauchy}
	Under the hypothesis of Theorem \ref{teo:cota-fn-gn-An}, there exists a constant $A$ such that
	$$
	|A_n-A|< \frac{L^{-n}}{2L^{1-\delta}(1-L^{-1})}\|f_0\|_q, \forall n \in \mathbb{Z}^+.
	$$
\end{cor}

Therefore, taking $\|f_0\|_q<\bar\epsilon$, with $\bar\epsilon$ given by Theorem \ref{teo:cota-fn-gn-An},
the previous results
togheter with
$R_{L^n,0}f_0 = R_{L,n-1} \circ \cdots \circ R_{L,1} \circ R_{L,0}f_0$, allow us to prove that
		\begin{equation}
		\label{cotafinal}
		\|L^{2n}u(L^n\cdot,L^{2n})-Af_1^*\|_q
		\leq \frac{C_{L,q,\delta} }{L^{n(1-\delta)}}\|f_0\|_q,
		\end{equation}
where $C_{L,q,\delta}=1+ \frac{k_q}{2L^{1-\delta}(1-L^{-1})}$.
From estimate (\ref{cotafinal}), we get that $\|tu(\sqrt{t}\cdot, t)
-Af_1^*\|_q \leq {C_{L,q,\delta}t^{-(1-\delta)/2}} \|f_0\|_q$
is valid for $t_n=L^{2n}$, $n \in \Z^+$, if $L>L_{\delta}$ and $f_0$ is sufficiently small. Furthermore, We can extend this bound to
$t = \tau L^n$, with $\tau \in [1,L]$ and $L > L_\delta$ by replacing
everywhere $L$ by $\tau^{1/n} L$, which finishes the proof of Theorem \ref{teo:lim-ass-burgers}.

\section{Generalization}
\label{sec:geral}

Consider the IVP
\begin{equation}
\label{pvi:n-lin-F-cap4}
\left\{
\begin{array}{ccccccc}
u_t = u_{xx} + \lambda F(u,u_x), \ x \in \mathbb{R}, \
t>1, \ \lambda \in [-1,1], \\
u(x, 1) = f(x), \ x \in \mathbb{R}, \ f(x) \in \B_q,
\end{array}
\right.
\end{equation}
where  $f(x)$ is now an odd function and $F(u,v)$ is analytical at $u=v=0$
and
\begin{equation}
	\label{def:n-linearid-F-cap4}
	F(u,v) = \sum_{m\geq a,\, n \geq b \atop 4a+3b-2>0}
	c_{m,n} u^{2m+1}v^{n}.
\end{equation}
with convergence radius $r= \min\{r_u, r_v\}>0$,
where $r_u$ and $r_v$ are the radii of convergence of the sums over $u$ and $v$, respectively.
Notice that, when $a=0$ and $b=1$ in (\ref{def:n-linearid-F-cap4}),
if $c_{m,n}=0$ for all $(m,n) \not= (0,1)$ and if $c_{0,1} = 1$ then we recover the Burgers equation (\ref{pvi:Eq-de-Burg}).

We shall prove that there exists a positive $\epsilon$ such that, if $f \in \B_q \cap B_\epsilon$, then the
solution to IVP (\ref{pvi:n-lin-F-cap4}) for $t \in[1, L^2]$, with $L>1$,
is given by $u(x,t) = u_f(x,t) + \lambda N(u)(x,t)$, where $u_f(x,t)$ is the solution to the linear IVP with initial data $f$
and $N(u)$ is given by
 \begin{equation}
N(u)(x,t) = \int_0^{t-1}
\left(
\int_{\mathbb{R}} \frac{e^{\frac{-(x-y)^2}{4s}}}{\sqrt{4 \pi s}}
F(u,u_x)(y, t-s-1)dy\right)ds,
\label{def:op-N-caso-geral}
\end{equation}
where $F(u,u_x)$ is the sum in (\ref{def:n-linearid-F-cap4}). Also notice that, if $F(u,u_x)$
is of the form $[h(u)]_x$, $h(u)$ being an analytic function at $ u = 0 $, then $\mathcal{F}\left\{[h(u)]_x\right\}(0, t) = 0$ and so
the solution $u(x,t)$ of IVP (\ref{pvi:n-lin-F-cap4}) in this case will
also satisfy the zero mass condition.

However, we are now considering that the power of the second variable of the representation (\ref{def:n-linearid-F-cap4}) can assume a value greater than 1 and this does not allow us to extract adequate (and necessary) information about the mass of the solution $u (x, t)$ as in the case above for disturbances of the type $[h(u)]_x$. In this sense, it is necessary that we restrict the set of initial data to odd functions in $\B_q \cap B_\epsilon$ and, to ensure that the
IVP (\ref{pvi:n-lin-F-cap4}) has the same parity as the initial data $f(x)$, we have included conditions on the powers of the sum (\ref{def:n-linearid-F-cap4}). Notice that if
$a$ is odd and
$b \in \Z^+$ then
any solution to $u_t=u_{xx}+\lambda u^au_{x}^b$
satisfies $u(x,t)=-u(-x,t)$
and this justifies the fact that the powers in the representation of $F(u,u_x)$ are chosen of the form $u^{2m+1}u_x^n$, with $m, n \in \Z^+$.

If we also consider that the initial data $f(x)$ of IVP (\ref{pvi:n-lin-F-cap4}) is odd, then the solution $u(x,t)$ to this problem will also be odd in the $x$
variable, for all $t \in [1, L^2]$. In particular, $\mathcal{F}\{F(u, u_x)\}(0, t) = 0$ for all $t \in [1, L^2]$.
		This property is essential
		and it allows us to complete step 4 of the RGA and, therefore, to close the first  iteration loop
		(and hence the {\it n-th}
		loop).

Assuming that the IVP (\ref{pvi:n-lin-F-cap4}) has a well defined solution $u(x,t)$
for all $t > 1$, considering the non-canonical scaling $L^2u(Lx,L^2t)$ and following the nomenclature introduced by
Bricmont et al. in \cite{bib:bric-kupa-lin}, it is clear that the restriction imposed to the sum (\ref{def:n-linearid-F-cap4}), that is, $m \geq a$ and $n \geq b$, with $a,b \in \Z^+$ satisfying $4a+3b-2>0$,
makes the nonlinearity $F(u,u_x)$ {\em irrelevant} in the RG sense.

In order to prove the existence and uniqueness of the solution to IVP (\ref{pvi:n-lin-F-cap4}) and therefore obtain Theorem \ref{teo:lim-ass-n-lin},
we need an upper bound for the Fourier Transform of the nonlinearity. The following proposition follows from the definition of the space
$B^{(L)}$ given by (\ref{def:esp-B}) and from the properties of the Fourier Transform.
\begin{prop}
	\label{lem:des-der-u-caso-ger}
	Given $q>3/2$ and $u \in B^{(L)}$:
	$$
	|\partial_{\omega}^i\widehat{u}(\omega,t)|, |\widehat{u_x}(\omega,t)|
	\leq
	\frac{2\|u\|}{1+|\omega|^{q-1}},
	\ i=0,1,2,
	$$
for all $\omega \in \R$ and $t > 1$.
\end{prop}
It is also important that, in this case,
		we make sure that the values
	 assumed
		by the solution $u(x, t)$ and its derivative $u_x (x, t)$ are within
		the analytic region of $F(u, u_x)$. To ensure this and other important results we will need the following:

\begin{prop}
	\label{prop:conv-F}
Given $u \in B_L$, there exists $K>0$ such that
$$
|u(x,t)|,|u_x(x,t)| < K\|u\|, \,\,\,\ \forall x \in \R \ \mbox{and} \ \forall t \in [1,L^2].
$$
\end{prop}
\proofname:
To prove the proposition we use the representation of the inverse Fourier transform for $u$ and $u_x$ and the definition of the $B_q$
space. Furthermore,
for $|\omega| \geq 1$, since $q>2$, we use that
$(|\omega|+|\omega|^{q})/({1+|\omega|^q})<2$.
Therefore, defining
\begin{equation}
\label{def:const-k}
K = \max \left\{
\frac{1}{2 \pi}\int_{\R}
\frac{1}{1+|\omega|^q}d \omega, \frac{1}{\pi}\int_{\R}
\frac{1}{1+|\omega|^{q-1}}d \omega
\right\}
\end{equation}
and using the Fourier representation of $u(x,t)$ and $u_x(x,t)$,
we get that
$
\left|u(x,t)\right|, \left|u_x(x,t)\right| \leq  K \|u\|.
$
\eop

{\bf Remark:}
		Notice that, if
		$f \in B_q$, $q>2$ and $u \in B_f$, it follows from the above proposition that
		\begin{eqnarray}
		\label{cota:desig-u-k-C}
		\left|u(x,t)\right|, \left|u_x(x,t)\right| \leq   K \bar{C}_L\|f\|_q,
		\end{eqnarray}
	where $\bar{C}_L$ is given in (\ref{des:oper-N}).

The proof of the next estimates, which we will use to prove Theorem \ref{teo-ex-uni-caso-geral}, are straightforward:
\begin{prop}
	\label{lema:cotas-caso-geral}
	If $q>1$ and $t \geq 1$, then
	\begin{equation}
	\frac{1}{1+|\omega|^{q-1}} \int_{0}^{t-1} e^{-\omega^2s}ds  <
		\frac{2t-1}{1+|\omega|^q} ,
		\label{ap:cota-F-t}
	\end{equation}
	
	\begin{equation}
		\frac{1}{1+|\omega|^{q-1}}\int_{0}^{t-1} |\omega|^i s e^{-\omega^2s}ds <
		\frac{t^2-2t+4}{1+|\omega|^q}, \ i = 0, \ 1,
		\label{ap:cota-M-t}
	\end{equation}
	and
	\begin{equation}
		\frac{1}{1+|\omega|^{q-1}}\int_{0}^{t-1}
		\omega^2s^2 e^{-\omega^2s}ds  <
		\frac{2(t-1)^3/3+2}{1+|\omega|^{q}},
		\label{ap:cota-T-t}
	\end{equation}
for all $\omega \in \R$.
	\end{prop}

\subsection{Local Existence and Uniqueness}
\label{sec:exis-uni-caso-geral}

Let the operator $T$, acting on functions
$u \in B_L$,  be defined by
$T(u)=u_f+\lambda N(u)$,
where $u_f(x,t)$ denotes the solution to the linear IVP with initial data $f(x)$
and $N(u)$ be given by (\ref{def:op-N-caso-geral}). Notice that, if we take $f$ such that $\|f\|_q<r\left(K\bar{C}_L\right)^{-1}$,
where $r=\min\{r_u,r_v\}$, then,
using (\ref{cota:desig-u-k-C}), $F(u,u_x)$ is well defined and so are the operators $T(u)$ and $N(u)$.
We shall prove the following:

\begin{teo}
\label{teo-ex-uni-caso-geral}
Suppose $F(u,v)$ is given by (\ref{def:n-linearid-F-cap4}), with $4a+3b-2>0$, and positive convergence radii
$r_u $ and $r_v$ and let $r=\min\{r_u,r_v\}$.
Given $q>2$, $L>1$ and $\lambda \in [-1,1]$, there exists $\epsilon = \epsilon (L, q, r, F)>0$	such that
if $f \in B_q$ and $\|f\|_q<\epsilon$, then there is a unique solution to IVP (\ref{pvi:n-lin-F-cap4}), for $t \in [1, L^2]$, in $B_f$.
\end{teo}
The proof of the above theorem is straightforward from the following two lemmas. The first one will guarantee that,
if the initial data is sufficiently small, then $T(B_f)\subset B_f$ and the second that
$T$ is a contraction in $B_f$. The unique fixed point of $T$ is then the unique solution to IVP (\ref{pvi:n-lin-F-cap4}) in $B_f$.
\begin{lema}
\label{lem-op-N-caso-geral}
Suppose $F(u,v)$ is given by (\ref{def:n-linearid-F-cap4}), with $4a+3b-2>0$, and positive convergence radii
$r_u $ and $r_v$ and let $r=\min\{r_u,r_v\}$.
Given $q>2$, $L>1$ and $\lambda \in [-1,1]$, there exists $\epsilon_1 = \epsilon_1(L, q, r, F)>0$	such that, if $f \in B_q$ and $\|f\|_q<\epsilon_1$,
then, for all $u \in B_f$,
$$
\|N(u)\| < \|f\|_q,
$$
where the operator $N$ is given by
(\ref{def:op-N-caso-geral})
.
\end{lema}
\proofname:
Given $q>2$, $f \in B_q$, $u \in B_f$ and $m,n \in \Z^+$ such that
$m \geq a$, $n \geq b$ and $4a+3b-2>0$, define, for $x \in \R$ and $t \in [1, L^2]$,
\begin{equation}
H_{m,n}(u)(x,t)=c_{m,n}
\int_0^{t-1}
\left(
\int_{\mathbb{R}} \frac{e^{\frac{-(x-y)^2}{4s}}}{\sqrt{4 \pi s}}
 [u^{2m+1}u_x^{n}](y,t-s-1)dy\right)ds
\label{def:operadot-H}
\end{equation}
and notice that
\begin{equation}
\label{eq:rep-op-N-caso-geral}
N(u)(x,t) = \sum_{m\geq a,\, n \geq b \atop 4a+3b-2>0} H_{m,n}(u)(x,t),
\end{equation}
if $\|f\|_q<r\left(K\bar{C}_L\right)^{-1}$.
Using the properties of the Fourier Transform, we can write $\widehat{H_{m,n}}(u)(\omega,t)$ as
{\small{
\begin{eqnarray*}
\frac{c_{m,n}}{(2\pi)^{2m+n}}\int_0^{t-1} e^{-\omega^2s} \int_{\mathbb{R}^{2m+n}}
\widehat{u}(\omega - p_1)
\cdots
\widehat{u}(p_{2m}-r_1)
\cdots
\widehat{u}_x(r_{n-1}-r_n)\widehat{u}_x(r_n)
 dpdrds,
 \label{eq:operador-H-Transf}
	\end{eqnarray*} }}
where integrand is evaluate at time $t-s-1$, with $2m$ convolutions of $\widehat{u}$ with $\widehat{u}$,
one convolution of $\widehat{u}$
with $\widehat{u}_x$ and $n-1$ convolutions of $\widehat{u}_x$ with
$\widehat{u}_x$. From Proposition \ref{lem:des-der-u-caso-ger},
$$
\int_{\mathbb{R}}
\left|\widehat{u}(\omega - p_1)
\right|\left|\widehat{u}(p_1)\right|
dp_1
\leq
2^2\|u\|^2\int_{\R}\frac{1}{1+|\omega-p_1|^{q-1}}
\cdot
\frac{1}{1+|p_1|^{q-1}}
dp_1.
$$
Since $q>2$, the integral in the right hand side of the inequality above is convergent and therefore,
$$
\left|	\widehat{H_{m,n}(u)}(\omega,t) \right| \leq
|c_{m,n}|
\frac{(4t-2)}{1+|\omega|^{q}}
\left( \frac{G_q}{\pi} \right)^{2m+n}
\|u\|^{2m+n+1},
\label{des:cota-H}
$$
where
\begin{equation}
G_q = \left(2^{q+1}+3\right)\int_{\R}\frac{1}{1+|x|^q}dx.
\label{eq:G-q}
\end{equation}
Using this estimate in (\ref{eq:rep-op-N-caso-geral}) we get
\begin{equation}
\left|	\widehat{N(u)}(\omega,t) \right| \leq
\frac{(4t-2)}{1+|\omega|^{q}}
\sum_{m\geq a,\, n \geq b \atop 4a+3b-2>0}
|c_{m,n}|
\left( \frac{G_q}{\pi} \right)^{2m+n}
\|u\|^{2m+n+1}.
\label{des:cota-N-caso-geral}
\end{equation}
We now define $r_0 = \min \{r/K, \pi r/G_q\}$ and take $\|f\|_q< \bar{C_L}^{-1}r_0$ to guarantee that the above sum is convergent.

In order to obtain an upper bound to
$\left|	\partial_{\omega} \widehat{N(u)}(\omega,t) \right|$, we first notice that in the definition of
$F(u,u_x)$ (see (\ref{def:n-linearid-F-cap4})), there are terms of the form $u^{2m+1}u_x^n$, with $m \geq a$ and $n \geq b$
such that $4a+3b-2>0$. Then, we use Propositions \ref{lem:des-der-u-caso-ger}, \ref{lema:cotas-caso-geral} and the fact that
given $q>1$,
$$
\int_{\R}\frac{1}{1+|x|^q}\cdot \frac{1}{1+|x-\omega|^q}
dx
\leq \frac{G_q}{1+|\omega|^q},
$$
for all $\omega \in \R$, where $G_q$ is given by (\ref{eq:G-q}),
to obtain:
$$\left|	\partial_{\omega}\widehat{N(u)}(\omega,t) \right| <
\frac{\left(4t^2-2t+14\right)}{1+|\omega|^{q}}
\sum_{m\geq a,\, n \geq b \atop 4a+3b-2>0}
|c_{m,n}|
\left( \frac{G_q}{\pi} \right)^{2m+n}
\|u\|^{2m+n+1}.
$$
and 
$$
\left|\partial_{\omega}^2 \widehat{N(u)}(\omega,t)\right|
<
\frac{(16t^3-12t^2-12t+170)}{1+|\omega|^{q}}
\sum_{m\geq a,\, n \geq b \atop 4a+3b-2>0}
|c_{m,n}|
\left( \frac{G_q}{\pi} \right)^{2m+n}
\|u\|^{2m+n+1},
$$
if $\|f\|_q< \bar{C_L}^{-1}r_0$. Using the above bounds we obtain, defining $C_{L}=16L^6-8L^4-10L^2+182$,
$$
\|N(u)\|  < C_L \sum_{m\geq a,\, n \geq b \atop 4a+3b-2>0}
|c_{m,n}|\left( \frac{G_q}{\pi} \right)^{2m+n}\|u\|^{2m+n+1}.
$$
Notice that, since $m \geq a$, $n \geq b$ and $4a+3b-2>0$, then $2m+n+1\geq 2$ and, if
$\|f\|_q< \bar{C_L}^{-1}r_0$, then
$\|u\| <r_0<r$ and
$$
\|u\|^2
\sum_{m\geq a,\, n \geq b \atop 4a+3b-2>0}
|c_{m,n}|
\left( \frac{G_q}{\pi} \right)^{2m+n}
\|u\|^{2m+n-1}
 <+ \infty.$$
 It follows that
$\|N(u)\|  < K_{L,q}\bar{C}_L^2\|f\|^2_q$, with
\begin{equation}
\label{eq:K-L-q}
K_{L,q} = C_L\sum_{m\geq a,\, n \geq b \atop 4a+3b-2>0}
|c_{m,n}|\left( \frac{G_q}{\pi} \right)^{2m+n}r_0^{2m+n-1}.
\end{equation}
Then, if $\|f\|_q< \epsilon_1$, defining $\epsilon_1 = \epsilon_1 (L,q,r,F)
\equiv
\min \left\{
(K_{L,q}\bar{C}_L^2)^{-1}, \bar{C}_L^{-1}r_0
\right\}$,
we obtain $\|N(u)\| < \|f\|_q$, for $u \in B_f$.
\eop

\begin{lema}
\label{lem:contracao-caso geral}
Under the hypothesis of Lemma \ref{lem-op-N-caso-geral}, there exists $\epsilon_2=\epsilon_2(L,q,r,F)>0$
such that, if $f \in B_q$ and $\|f\|_q <\epsilon_2$, then
$$
\|N(u)-N(v)\| < \frac{1}{2}\|u-v\|, \,\,\ \forall u,v \in B_f.
$$
\end{lema}
\proofname:
Given $q>2$ and $f \in B_q$, let $u, v \in B_f$ and $H_{m,n}(u)$ given by (\ref{def:operadot-H}), where
$m,n \in \Z^+$ are such that $m \geq a$, $n \geq b$ with $4a+3b-2>0$. Then, we can write
$\left[\widehat{H_{m,n}(u)} - \widehat{H_{m,n}(v)}\right](\omega,t) = I_1 + I_2$, where, for $i=1,2$,
$$
I_j = \frac{c_{m,n}}{(2\pi)^{2m+n}}\int_0^{t-1}e^{-\omega^2s}F_j(\omega,s)ds,
$$
$F_1(\omega,s)=(\widehat{u}-\widehat{v})*\widehat{u}* \cdots *\widehat{u}*\widehat{u_x}* \cdots*\widehat{u_x}$ and
$F_2(\omega,s)= \widehat{v}*\widehat{u}* \cdots*\widehat{u}*\widehat{u_x}* \cdots*\widehat{u_x}
-\widehat{v}* \cdots*\widehat{v}*\widehat{v_x}* \cdots*\widehat{v_x}$.
Proceeding as in Lemma \ref{lem-op-N-caso-geral}, we get
$$
|I_1| <
|c_{m,n}|
\frac{(4t-2)}{1+|\omega|^{q}}
\left( \frac{G_q}{\pi} \right)^{2m+n}
\|u-v\|
\|u\|^{2m+n},
$$
where $G_q$ is the contant given by (\ref{eq:G-q}). In order to estimate $|I_2|$ we sum and subtract
$
\widehat{v}*\widehat{v}
*\widehat{u}
*\cdots
*\widehat{u}*\widehat{u_x}
*\cdots
*\widehat{u_x}
$
in the integrand of $I_2$,
and obtain to other integrals. The first one which can be bound by
$
|c_{m,n}|
\frac{(4t-2)}{1+|\omega|^{q}}
\left( \frac{G_q}{\pi} \right)^{2m+n}
\|u-v\|
\|u\|^{2m+n-1}\|v\|$ and the second can be similarly decomposed in two other integrals by summing and subtracting
$
\widehat{v}*\widehat{v}
*\widehat{v}*\widehat{u}
*\cdots
*\widehat{u}*\widehat{u_x}
*\cdots
*\widehat{u_x}
$
in its integrand.
Using that
$|\widehat{u_x}(\omega,t)-\widehat{v_x}(\omega,t)|
\leq 2\|\widehat{u}-\widehat{v}\|
(1+|\omega|^{q-1})^{-1}$,
see Proposition \ref{lem:des-der-u-caso-ger}, and repeting this procedure
$2m+n-3$ times, we obtain, as an upper bound for $\left| \left[
\widehat{N (u)} - \widehat{N (v)}\right](\omega, t)\right|$,
$$
\frac{(4t-2)}{1+|\omega|^q}
\|u-v\|
\sum_{m\geq a,\, n \geq b \atop 4a+3b-2>0}
\left[
|c_{m,n}|
\left( \frac{G_q}{\pi} \right)^{2m+n}
\left(
\sum_{i=0}^{2m+n}\|u\|^{2m+n-i}\|v\|^i
\right)\right].
$$
Notice that, since $\|f\|_q< \bar{C_L}^{-1}r_0$ and $u,v \in B_f$, the sum in the right hand side of
the inequality above is convergent.
Similarly, we get the bounds for $\left| \partial_\omega
\left[\widehat{N (u)}- \widehat{N (v)}\right](\omega, t)\right|$ and
$\left| \partial^2_\omega\left[\widehat{N (u)}- \widehat{N (v)}\right](\omega, t)\right|$, respectively:
\begin{eqnarray*}
\frac{\left(4t^2-2t+14\right)}{1+|\omega|^q}
\|u-v\|
\sum_{m\geq a,\, n \geq b \atop 4a+3b-2>0}
|c_{m,n}|
\left( \frac{G_q}{\pi} \right)^{2m+n}
\sum_{i=0}^{2m+n}\|u\|^{2m+n-i}\|v\|^i,
\end{eqnarray*}
and
\begin{eqnarray*}
\frac{(16t^3-12t^2-12t+170)}{1+|\omega|^q}
\|u-v\|
\sum_{m\geq a,\, n \geq b \atop 4a+3b-2>0}
|c_{m,n}|
\left( \frac{G_q}{\pi} \right)^{2m+n}
\sum_{i=0}^{2m+n}\|u\|^{2m+n-i}\|v\|^i,
\end{eqnarray*}
It follows that $\|N (u)-N (v)\| <\bar{K}_{L,q}\|f\|_q\|u-v\|$, where
$$
\bar{K}_{L,q}=C_L\bar{C}_L
\sum_{m\geq a,\, n \geq b \atop 4a+3b-2>0}
(2m+n+1)
|c_{m,n}|
\left( \frac{G_q}{\pi} \right)^{2m+n}
r_0^{2m+n-1}
$$
and, defining $\epsilon_2 =
\epsilon_2(L,q,r,F)
\equiv \min \left\{(2\bar{K}_{L,q})^{-1},\bar{C}_L^{-1}r_0\right\}$ and taking $\|f\|_q < \epsilon_2$, we prove the lemma.
\eop

\subsection{Renormalization and the proof of Theorem \ref{teo:lim-ass-n-lin}}
\label{sec:ind-caso-geral}

The algorithm described in Section \ref{sec:arg} also applies in more general cases, such as IVP (\ref{pvi:n-lin-F-cap4}) with odd initial
data $f$ (in particular, zero mean initial data). From now on we consider $L>L_0$, where $L_0$
is given by the Contraction Lemma \ref{lema:contract}. Furthermore, whenever we refer to the ball $B_\epsilon$,
given by (\ref{def:bola-Bf}), we are considering $\epsilon$ given in Theorem \ref{teo-ex-uni-caso-geral}.

Given $q>2$ and $\lambda \in [-1,1]$, for
$k \in \{0, 1, 2, \cdots\}$,
let us consider IVP $(PVI_{k}, f_{k})$, the k-th iteration of the RGA,
\begin{equation}
\left\{
\begin{array}{ccccccc}
u_t = u_{xx} + \lambda_k F_{L,k}(u,u_x), \ x \in \mathbb{R}, \ t \in (1, L^2], \\
u(x, 1) = f_k(x), \ f_k \in B_q,
\label{pvi:n-lin-F-L-k}
\end{array}
\right.
\end{equation}
where $\lambda_k =\lambda L^{-k(4a+3b-2)}$,
$$
F_{L,k}(u,u_x)= \sum_{m\geq a,\, n\geq b \atop 4a+3b-2>0}c_{m,n} L^{k[4(a-m)+3(b-n)]}u^{2m+1}u_x^{n},
$$
and $f_k \in B_\epsilon$ is odd and will be defined next.
Taking $\epsilon = \min\{\epsilon_1, \epsilon_2\}$, where $\epsilon_1$ and $\epsilon_2$
were obtained, respectively, from lemmas 4.1 and 4.2, we have that
$\|f_k\|_q < \epsilon \leq \bar{C}_L^{-1}r_0$ and
$L^{k[4(a-m)+3(b-n)]}\leq 1$. Therefore, from Theorem \ref{teo-ex-uni-caso-geral}, it follows that $F_{L,k}(u, u_x)$ is well defined
and that there is a unique local solution
to the IVP (\ref{pvi:n-lin-F-L-k}) in $B_{f_k}$ given by
\begin{equation}
\label{eq:sol-u-L-k}
u_k(x,t) = u_{f_k}(x,t) + \nu_k(x, t)
\end{equation}
where $u_{f_k}$ is the solution to the linear IVP associated, with initial data $f_k$ and $\nu_k = \lambda_k N(u_k)$,
with $N(u_k)$ given by (\ref{def:op-N-caso-geral}).
For each $k \in \{0, 1, 2, \cdots\}$, we define
\begin{equation}
\label{def:rg-n-lin-caso-ger}
R_{L,k}f_k(x) \equiv L^2u_k(L x,L^2),
\
\forall x \in \R.
\end{equation}
and
\begin{equation}
\label{def:fk-caso-geral}
f_0 (x) = f(x) \ \mbox{e} \
f_{k+1}(x)= R_{L,k}f_k(x) ,
\ \forall x \in \R.
\end{equation}
In the sequel, we state a version of the nonlinear Renormalization Lemma (see Lemma \ref{lema:renormal}) which
is suitable for implementing the induction step that allows iterating the RGA.
\begin{lema}
		\label{lema:renormal-caso-geral}
	Consider IVP (\ref{pvi:n-lin-F-L-k}), with $t \in [1,  L^2]$ and initial data $f_k \in B_\epsilon$
		such that $f_k$ admits the decomposition
		\begin{equation}
		\label{eq:decomp-fk}
		f_{k}=A_{k}f_1^*+g_{k},
		\end{equation}
where $A_k$ is a constant, $f_1^*$ is given by (\ref{def:pont-fixo}),
and $g_k \in B_q$ such that $\widehat{g_k}(0)=\widehat{g_k}'(0)=0$. Then:
		\begin{enumerate}
	\item[ (a)] $f_{k+1}$, given by (\ref{def:fk-caso-geral}), can be decomposed as
$f_{k+1}=A_{k+1}f_1^*+g_{k+1}$, where $A_{k+1}=A_k-i\widehat{\nu_k}'(0),$
$g_{k+1} = R_Lg_k + L^2\nu_k(L\cdot) +i\widehat{\nu_k}'(0)f^*_1$
			and $\nu_k = \lambda_kN(u_k)$. Furthermore, $g_{k+1}\in B_q$ is such that
			$\widehat{g_{k+1}}(0)=\widehat{g_{k+1}}'(0)=0$. In particular, $f_{k+1}$ is odd.
	\item[ (b)] $|A_{k+1} - A_k| \leq |\lambda_k|K_{L,q}\bar{C_L}^2\|f_{k}\|_q^2$,
		where $\bar{C_L}$ e $K_{L,q}$ are given, respectively, by (\ref{eq:C-barra}) and (\ref{eq:K-L-q}).
  \item[ (c)] $\|g_{k+1}\|_q \leq \frac{C}{L}\|g_{k}\|_q+		|\lambda_k|\bar{E}_{L,q}\|f_{k}\|_q^2$,
			where $C$ is the constant given in the Contraction Lemma \ref{lema:contract} and
			\begin{equation}
			\bar{E}_{L,q}=(k_q+L^{q+1}) K_{L,q}\bar{C_L}^2,
			\label{eq:E-L-q-barra}
			\end{equation}
		with $k_q$ given by (\ref{des:cota-pont-fixo}).
		\end{enumerate}
	\end{lema}
\proofname: Given $k \in \Z^+$, suppose that the initial data $f_k$ of the IVP
(\ref{pvi:n-lin-F-L-k}) is an odd function of $B_q \cap B_\epsilon$, $q>2$.
It follows from Theorem \ref{teo-ex-uni-caso-geral} that there is a unique solution
$u_{k}(x,t)$ in $B_{f_{k}}$ to IVP (\ref{pvi:n-lin-F-L-k}) with initial data $f_{k}$ and also
$u_{k}(x,t)$ is an odd funtion in the $x$ variable. Therefore, if  $f_{k+1}$ is given by (\ref{def:fk-caso-geral}),
using definition (\ref{def:rg-n-lin-caso-ger}) we get that $f_{k+1}$ is also odd.
The rest of the proof follows as in \cite{bib:braga-furt-mor-rolla-tp}.
\eop

From now on we assume that $L>L_\delta$ and define $D_1$ as in (\ref{D1geral}) with $G_{L,q}$ replaced by $K_{L,q}\bar{C}_L^2$ and, for
$k=1,2,\cdots$, we define $D_{k+1}$ as
\begin{eqnarray}
\label{def:D-k-caso-geral}
 \frac{1}{L^{(k+1)(1-\delta)}}
 +
 k_q\left(1+K_{L,q}\bar{C}_L^2\|f_0\|_q+
 K_{L,q}\bar{C}_L^2\|f_0\|_q
 \sum_{j=1}^{k}
 \frac{D_j^2}{L^{j(4a+3b-2)}}
 \right),
\end{eqnarray}
where $k_q$, $\bar{C}_L$ and $K_{L,q}$ are given, respectively, by (\ref{des:cota-pont-fixo}),
(\ref{eq:C-barra}) and (\ref{eq:K-L-q}), and $4a+3b-2>0$. Notice that, if
$\|f_0\|_q<\frac{1}{2L^{1-\delta}\bar{E}_{L,q}D^2}$, where $\bar{E}_{L,q}$ is given by (\ref{eq:E-L-q-barra}),
we can show that $D_k<D$, for all $k \in \Z^+$, with $D$ and $D_k$ given, respectively by (\ref{eq:D}) and (\ref{def:D-k-caso-geral}).

If $\epsilon>0$ is the one given in Theorem \ref{teo-ex-uni-caso-geral} and the initial data $f=f_0$ is such that $\|f_0\|_q<\epsilon$, then we can iterate the procedure and, at each step of the algorithm we must guarantee that the initial data is suficiently small. In this case, since
$L^{k(4a+3b-2)(1-\delta)}>L^{k(1-\delta)}$, in the $k$-th iteration of the RGA, if $\|f_0\|_q<\bar\epsilon_{k}$, then $f_k \in B_{\epsilon}$, where
$$
\bar{\epsilon}_0 \equiv \epsilon \ \mbox{and} \
\bar{\epsilon}_{k+1} = \min \left\{
\frac{1}{2L^{k(1-\delta)}\bar{E}_{L,q}D_k^2}, \bar{\epsilon}_{k}, \frac{\epsilon}{D_{k+1}}
\right\}, \ \forall k \in \Z^+,
$$

Defining
\begin{equation}
\label{def:epsilon-barra-cap4}
\bar{\epsilon} = \min \left\{
\frac{1}{2L^{1-\delta}\bar{E}_{L,q}D^2}, \frac{\epsilon}{D}
\right\},
\end{equation}
we can prove the following
\begin{teo}
	\label{teo:cota-fn-gn-caso-ger}
	Given $\delta \in (0,1)$, $a,b \in \Z^+$ such that $4a+3b-2>0$ and $L>L_{\delta}$, with $L_{\delta}$ given by (\ref{eq:L-delta}),
	there exists $\bar{\epsilon}>0$ such that, if $\|f_0\|_q<\bar{\epsilon}$ and $f_0$ is odd, then, for all $k=0, 1, 2, \cdots$, $f_{k+1}$
	given by (\ref{def:fk-caso-geral}) and (\ref{def:rg-n-lin-caso-ger}) is well defined, odd and can be written as (\ref{eq:decomp-fk}), where
	$\hat{g_k}(0)=\hat{g_k}'(0)=0$,
	\begin{equation}
	\|g_k\|_q
	\leq
	\frac{1}{L^{k(1-\delta)}}\|f_0\|_q.
	\label{des:decaim-gk-caso-ger}
	\end{equation}
	Furthermore,
	\begin{equation}
	\|f_k\|_q \leq D_k\|f_0\|_q,
	\label{des:cota-fk-caso-geral}
	\end{equation}
	with $D_k$ given by (\ref{def:D-k-caso-geral}) and, in particular, $\|f_k\|_q < \epsilon$.
\end{teo}
With Lemma \ref{lema:renormal-caso-geral} and Theorem \ref{teo:cota-fn-gn-caso-ger},
the $k$-th iteration of the RGA is complete for all
$k \in \{ \ 0, \ 1, \ 2, \cdots\}$ and we are able to finally prove Theorem \ref{teo:lim-ass-n-lin}.

{\bf Proof of Theorem \ref{teo:lim-ass-n-lin}:}
First, we observe that the nonlinear RG operator $R_{L,k}$ defined in (\ref{def:fk-caso-geral})
satisfies the semigroup property, that is,
$R_{L^k,0} = R_{L,k-1} \circ \cdots \circ R_{L,1} \circ R_{L,0}$, for every $k \in \{1, 2, \cdots \}$.
Now, since $\|f_0\|_q< \bar\epsilon$ and $f_0$ is odd, it follows from Theorem \ref{teo:cota-fn-gn-caso-ger}
that $f_k=A_kf_1^*+g_k$ for all $k=1, 2, \cdots$,
where $A_k$ is constant, that $g_k \in B_q$, $q>2$, has both zero mass and zero first moment, and that the
inequality (\ref{des:decaim-gk-caso-ger}) is valid.
From (\ref{def:fk-caso-geral}) and the semigroup property,
$$
\|L^{2k}u(L^k\cdot, L^{2k}) - A_kf_1^*\|_q =
 \|g_k\|_q
\leq \frac{\|f_0\|}{L^{k(1-\delta)}}
$$
and the $A_k$'s satisfy
$$
|A_k-\bar{A}|\leq \frac{L^{-k(4a+3b-2)}}{2L^{1-\delta}(1-L^{-(4a+3b-2)})}\|f_0\|_q,
$$
where $\bar{A}$ is a constant depending on the initial data $f$ and on the nonlinearity $F$.
Since $\delta \in (0,1)$, $L>1$ and $4a+3b-2>0$, it follows that $L^{1-\delta}L^{-(4a+3b-2)}<1$ and, since
$\|f_1^*\|_q<k_q$, with $k_q$ given by (\ref{des:cota-pont-fixo}), using the triangle inequality and the two previous inequalities, we get:
$$
\label{des:comp-ass-u-n}
\|L^{2k}u(L^{k}\cdot, L^{2k})
-\bar{A}f_1^*\|_q
\leq \frac{\bar{C}}{L^{k(1-\delta)}}\|f_0\|_q,
\ \forall k \in \{0,1, \cdots \},
$$
where
$\bar{C} = \bar{C}(L,q,a,b,\delta)
\equiv 1 + k_q\frac{1}{2L^{(1-\delta)}\left(1-
L^{-(4a+3b-2)}\right)}$.
To conclude this demonstration, it is enough to notice that the above inequality is valid for
all $t_k=L^{2k}$, with $k \in \{0, 1, 2, \cdots \}$ and $L>L_\delta$, and we can extend
it to all $t \geq 1$ by taking $t = \tau L^{2k_0}$, $\tau \in (1, L^2)$, and $k_0 \in \Z^+$. Therefore,
$$
\|tu(\sqrt{t} \ \cdot, t)
-\bar{A}f_1^*\|_q
\leq \frac{\bar{C}}{t^{(1-\delta)/2}}\|f_0\|_q
$$
and, taking the limit $t\to \infty$ on both sides of the above inequality, we conclude the proof.
\eop

{\bf Acknowledgments:}

\clearpage
\parskip 0pt
\baselineskip = 18pt


\end{document}